\documentclass[12pt,english]{article} 

\usepackage{babel}
\usepackage{epsfig} 

\newcommand{\beq}{\begin{equation}}
\newcommand{\eeq}{\end{equation}}
\newcommand{\bea}{\begin{eqnarray}}
\newcommand{\eea}{\end{eqnarray}}
\newcommand{\lc}{\Lambda_{\scriptscriptstyle c}^+}
\newcommand{\alc}{\Lambda_{\scriptscriptstyle c}^-}
\newcommand{\lcc}{\Lambda_{\scriptscriptstyle c}^\pm}
\newcommand{\lccc}{\Lambda_{\scriptscriptstyle c}}
\newcommand{\xf}{x_{\scriptscriptstyle F}}
\newcommand{\pt}{p_{\scriptscriptstyle T}^2}
\newcommand{\xu}{x_{\scriptstyle u}}
\newcommand{\xd}{x_{\scriptstyle d}}
\newcommand{\xc}{x_{\scriptstyle c}}
\newcommand{\xup}{x_{\scriptstyle u'}}
\newcommand{\xcb}{x_{\scriptstyle {\bar c}}}
\newcommand{\xub}{x_{\scriptstyle {\bar u}}}
\newcommand{\xdb}{x_{\scriptstyle {\bar d}}}
\newcommand{\xii}{x_{\scriptstyle i}}

\newcommand\cpc[3]{{ Comput.\ Phys.\ Commun.\ }{\bf #1} (#2) #3}

\newcommand\epjc[3]{{ Eur.\ Phys.\ J. }{\bf C#1} (#2) #3}

\newcommand\npb[3]{{ Nucl.\ Phys.\ }{\bf B#1} (#2) #3}

\newcommand\plb[3]{{ Phys.\ Lett.\ }{\bf B#1} (#2) #3}

\newcommand\prd[3]{{ Phys.\ Rev.\ }{\bf D#1} (#2) #3}

\newcommand\prep[3]{{ Phys.\ Rep.\ }{\bf #1} (#2) #3}
\newcommand\prl[3]{{ Phys.\ Rev.\ Lett.\ }{\bf #1} (#2) #3}

\newcommand\zpc[3]{{ Z.\ Physik }{\bf C#1} (#2) #3}

\newcommand\sjnp[3]{{ Sov.\ J.\ Nucl.\ Phys.\ }{\bf #1} (#2) #3}

\newcommand\ibid[3]{{\it ibid.\ }{\bf #1} (#2) #3}
\newcommand{\etal}{{\it et al. }}

\newcommand{\hepex}[1]{{\tt hep-ex/#1}}

\begin{document}

\title{\bf $\lc-\alc$ production asymmetries in two component models}

\author{{\bf J. Magnin}\thanks{jmagnin@uniandes.edu.co}\\
{\small and}\\
{\bf L.M. Mendoza Navas} \\
{\normalsize {\it Depto. de F\'{\i}sica, Universidad de los Andes}},\\
{\normalsize {\it AA 4976, Santaf\'e de Bogot\'a, Colombia}}}

\date{}

\maketitle

\begin{abstract}
Experiments on charm hadroproduction have shown a substantial 
difference in the production of charm and anticharm hadrons. 
In this work we study $\lc$ and $\alc$ inclusive production in $p-N$ 
interactions in the framework of two component models.
We show that the recombination two component model 
gives a qualitatively and quantitatively good description of 
$\lcc$ production while the intrinsic charm model seems to be 
ruled out by recent experimental data from the SELEX Collaboration.
\end{abstract}


It is well known from experiments that there is a substantial difference in 
the production of charm and anticharm hadrons in hadron-hadron 
interactions. Leading (L) particles, which share one 
or more valence quarks with the initial hadrons, are favored in the incident 
hadron direction  over Non-Leading (NL) particles, which share none. 
This effect, known as Leading Particle Effect, has been observed 
in  inclusive production of charm mesons and baryons in $\pi^--N$,  
$p-N$, $K-N$ and $\Sigma^--N$ interactions by 
several experiments~\cite{others,e791,selex} . 

Leading particle effects can be quantified by means of a 
production asymmetry, which is defined as
\beq
A\equiv \frac{d\sigma^L-d\sigma^{NL}}{d\sigma^L+d\sigma^{NL}} \; .
\label{eq0}
\eeq

Asymmetries in charm-anticharm production indicate that charm 
hadronization cannot proceed by independent fra\-gmentation alone. 
They imply also that some sort of recombination mechanism, involving 
valence quarks in the initial hadrons, must take place in the production.

Several models have been proposed to explain 
charm hadroproduction. However, no theoretical consensus has been 
reached yet on what is the dominant mechanism giving rise to the 
observed asymmetries. This is partly due to the lack of 
simultaneous measurements of charm-an\-ti\-charm asymmetries and 
inclusive particle distributions in the 
same experiment. Actually, as models depend on a set of parameters, and 
these parameters can be adjusted to describe either the 
charm-anticharm asymmetries or the inclusive particle distributions, a 
meaningful comparison of models to experimental data must be done 
on both, asymmetries and inclusive particle distributions. 

Among the proposed models are the String Fragmentation 
model (SF)~\cite{lund}, implemented in the Lund 
Pythia-Jetset package~\cite{pythia}; the recombination of charm 
quarks produced perturbatively with the remnants of the 
initial hadrons~\cite{russian}; the Intrinsic Charm model (IC)~\cite{ic} 
and the recombination two component model (R2C)~\cite{nos-prd}. 

Aiming to extract information on the charm hadron production mechanisms, 
in this work we shall compare predictions of both, the IC and R2C two 
component models to recent experimental data on $p+N\rightarrow\lcc+X$ 
by the SELEX Collaboration~\cite{selex}.

In two component models the total cross section for charm hadron 
production receives contributions from two different processes, 
namely perturbative production of a $c\bar{c}$ pair in QCD followed 
by independent fragmentation, and contributions coming from another,  
non-perturbative, mechanism. So far, two possibilities have been 
considered for this second contribution: IC coalescence~\cite{ic} and 
the recombination of charm quarks, already present in the 
sea of the initial hadrons, with valence and sea quarks from the 
initial hadrons~\cite{nos-prd}. Then, in the $p+N\rightarrow \lcc + X$ 
reaction, 
\beq
\frac{d\sigma}{d\xf} = \frac{d\sigma^{Frag.}}{d\xf} 
+ \frac{d\sigma^{IC(Rec.)}}{d\xf} \; .
\label{eq1}
\eeq
The first term in the RHS of Eq.~(\ref{eq1}) describes the production 
of charm hadrons by independent fragmentation, while the second one 
is the contribution to $\lcc$ inclusive production coming from the IC 
or the recombination mechanisms.

Charm hadron production by independent fragmentation is given by
\bea
\frac{d\sigma^{Frag.} }{d\xf}&=&\frac{1}{2} \sqrt{s} \int H^{ab}
(x_a,x_b,Q^2) \nonumber \\
&\times&\frac{1}{E} \frac{D_{\lccc} \left( z \right)}{z} 
dz d\pt dy \: ,
\label{eq2} 
\eea
where 
\beq
D_{\lccc/c}(z) = \frac{N_{\lccc}}{z\left[1-1/z-\epsilon_c/(1-z)\right]^2}
\label{eq2b}
\eeq
is the Peterson fragmentation function~\cite{peterson} with $N_{\lccc}$ 
a normalization constant. $H^{ab}(x_a,x_b,Q^2)$ contains information on 
the initial hadron structures and the dynamics of the perturbative QCD 
(pQCD) charm production. At Leading Order (LO) it is given by
\bea
H^{ab}(x_a,&x_b&,Q^2) =  \Sigma_{a,b} \left[ q_a(x_a,Q^2)
\bar{q_b}(x_b,Q^2)\right. \nonumber \\
                   & + & \left. \bar{q_a}(x_a,Q^2) q_b(x_b,Q^2) \right]
\frac{d \hat{\sigma}}{d \hat{t}} \left|_{q\bar{q}} \right. \nonumber \\
                   & + & g_a(x_a,Q^2) g_b(x_b,Q^2) \frac{d \hat{\sigma}}
{d \hat{t}}\left|_{gg}\right. \;.
\label{eq3}
\eea
In Eqs.~(\ref{eq2}) (\ref{eq2b}) and (\ref{eq3}), $\xii$; $i=a,b$; 
is the momentum fraction of the initial hadron carried by the 
parton $i$, 
$z$ and $\pt$ are the momentum fraction of the initial hadron carried 
by the charm quark and its transverse momentum squared respectively, 
and $y$ is the rapidity of the charm antiquark. The sum in 
Eq.~(\ref{eq3}) runs over light and strange quarks. 

\begin{figure}[ht]
\centerline{\epsfig{figure=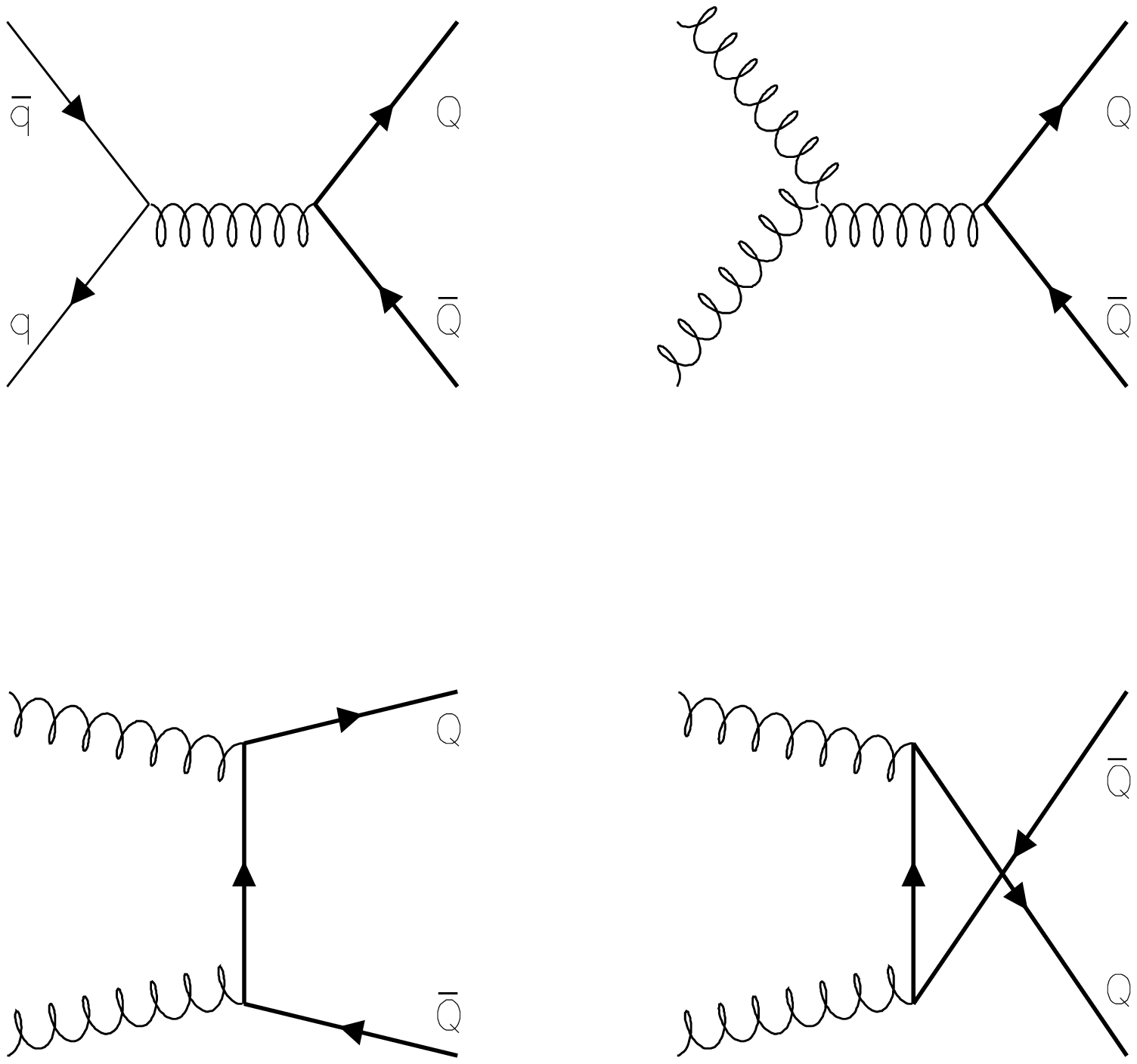,width=3.5in}}
\caption{Parton fusion processes contributing to $c\bar{c}$ 
perturbative production at LO.}
\label{fig0}
\end{figure}

Up to LO in the pQCD processes (see Fig.~\ref{fig0}), no contribution to 
the $\lc-\alc$ asymmetry arises from the charm quark production. 
At Next to Leading Order (NLO), a small $c-\bar{c}$ asymmetry 
translates into a tiny $\lc-\alc$ asymmetry~\cite{nason}. 
However, this effect is very small and has the opposite sign to 
the experimentally observed asymmetry in $p+N$ interactions. 
Then, the second term in the RHS of Eq.~(\ref{eq1}) must give the dominant 
contribution to the $\lc-\alc$ asymmetry. 

In the IC model for the $p+N\rightarrow \lcc+X$ reaction in the $\xf>0$ region, 
the second term in the RHS of Eq.~(\ref{eq1})
comes from fluctuations of protons in the beam to 
$\left|uudc\bar{c}\right>$ Fock states~\cite{ic}. These Fock states break up in 
the collision contributing to $\lc$ production through the coalescence of the 
intrinsic charm quark with $u$ and $d$ quarks. 
To obtain a $\alc$ purely from this process, a fluctuation of the 
proton to a $\left|uudu\bar{u}d\bar{d}c\bar{c}\right>$ Fock state is 
required. As the pro\-ba\-bi\-lity of the later is smaller than for the former, 
$\lc$ production is favored in the proton direction. 
The $\lc$ differential cross section for the intrinsic charm 
process is~\cite{ic}
\bea 
\frac{d\sigma^{IC} }{d\xf} & = & r^{IC} 
\int_0^1d\xu d\xup d\xd d\xc d\xcb \nonumber \\
\times  &\delta& \left( \xf-\xu-\xd-\xc\right)
\frac{dP^{IC}}{d\xu ...d\xcb} \; ,
\label{eq4} 
\eea 
where
\bea
\frac{dP^{IC}}{d\xu ...d\xcb} &=& N_5 \alpha_s^4\left(M_{c
\bar{c}}^2\right) \nonumber \\
&\times &\frac{\delta \left(1-\Sigma _{i=u}^{\bar c} x_i
\right)}{\left( m_p^2 - \Sigma _{i=u}^{\bar c} \hat {m}_i^2 x_i \right)^2}
\label{eq5} 
\eea
is the probability of the $\left|uudc\bar{c}\right>$ fluctuation of the 
proton, and $r^{IC}$ is a parameter which must be fixed from 
experimental data. Neglecting contributions coming from the 
$\left|uudu\bar{u}d\bar{d}c\bar{c}\right>$ Fock states of the proton, 
which are very small, the $\alc$ differential cross section is 
given only by the first term in the RHS of Eq.~(\ref{eq1}).

In the R2C model, the second term in the RHS of 
of Eq.~(\ref{eq1}) has the form~\cite{nos-prd,ranft}
\begin{eqnarray}  
\frac{d\sigma^{Rec.}}{d\xf}&=& r^{Rec}
\int_0^{1}\frac{d\xu}{\xu}\frac{d\xd}{\xd}
\frac{d\xc}{\xc}F_3\left( \xu,\xd,\xc\right) \nonumber \\
&\times& R_3\left( \xu,\xd,\xc,\xf\right) \;,
\label{eq6} 
\end{eqnarray} 
where $F_3$ and $R_3$ are the multiquark distribution and the 
recombination functions respectively. As for the intrinsic charm 
model, $r^{Rec}$ is a parameter which must be fixed from 
experimental data. 

For $\lc$ production in $p+N$ interactions in the $\xf > 0$ region, 
$F_3$ is given by
\bea
F_3\left( \xu,\xd,\xc\right) &\sim&
\xu u(\xu )\xd d(\xd )\xc c(xc )\nonumber \\
&\times& \rho(\xu,\xd,\xc)\; ,
\label{eq7}
\eea
while for $\alc$ production
\bea
F_3\left( \xub,\xdb,\xcb\right) &\sim&
\xub\bar{u}(\xub)\xdb\bar{d}(\xdb)\xcb\bar{c}(\xcb)\nonumber \\
&\times& \rho(\xub,\xdb,\xcb)\; .
\label{eq8}
\eea
In Eqs.~(\ref{eq7}) and (\ref{eq8}), $xq(x)$ is the q-flavored quark 
distribution in the proton. Notice that the multiquark distribution of 
Eq.~(\ref{eq7}) receives contributions from $u$ and $d$ valence and sea quarks, while 
Eq.~(\ref{eq8}) is constructed only by using sea quark distributions. Thus, 
$\alc$ production is due solely to the recombination of antiquarks popped up 
from the vacuum in the interaction. As quarks and antiquarks 
are created in pairs from the va\-cuum, $\lc$'s can also be formed by the 
recombination of $u$ and $d$ sea quarks, in addition to the 
$u$ and $d$ valence quarks, with charm quarks, thus giving rise to a 
$\lc-\alc$ asymmetry.

The momentum correlation function $\rho$, which we assume to be the same 
for both $\lc$ and $\alc$ production, is usualy taken as~\cite{ranft}
\beq
\rho(\xu,\xd,\xc)=(1-\xu -\xd -\xc)^\gamma \; ,
\label{eq8a}
\eeq
with the exponent $\gamma$ fixed appealing to some consistency 
condition like~\cite{ranft}
\bea
xq\left(x_i\right) & = & \int_0^{1-x_i}dx_j \nonumber \\
&\times&\int_0^{1-x_i-x_j}dx_k \:F_3 \left( \xu,\xup,\xd \right) \nonumber \\
                      &   &i,j,k = u,u',d 
\label{eq8b} 
\eea
for the valence quarks in the proton.

\begin{figure}[ht]
\centerline{\epsfig{figure=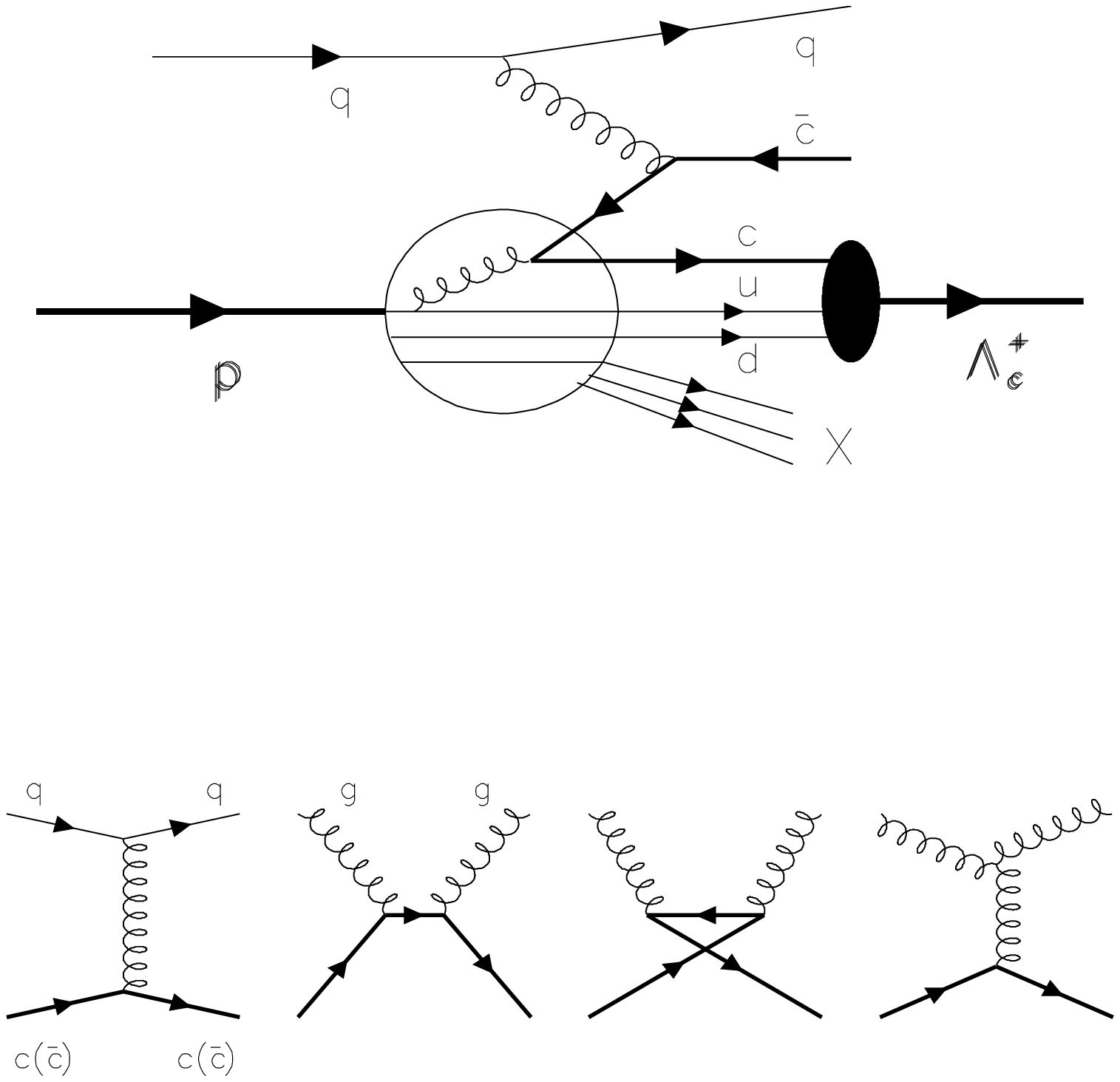,width=3.5in}}
\caption{Upper: $\lc$ production by recombination of a $c$ quark in the 
sea of the proton with $u$ and $d$ quarks. Lower: flavor exitation 
diagrams contri\-bu\-ting to $\lc$ and $\alc$ production by recombination.}
\label{fig0b}
\end{figure}

In Eqs.~(\ref{eq7}) and (\ref{eq8}) we are assuming the existence of 
charm quarks inside the proton. Indeed, assuming that the global scale 
of the whole process is of the order of $Q^2\sim 4m_c^2$, there should 
be a substantial contribution of charm quarks to the 
proton structure~\cite{grv}. However, charm quarks in the proton 
may have a twofold origin, namely, 
a non-perturbative component which must exist over a time scale 
independent of $Q^2$, and a perturbative component due to the QCD 
evolution. The non-perturbative contribution is expected to be 
small, of the order of $1\;\%$ or less~\cite{brodsky0}, and at 
$Q^2\sim 4m_c^2$ the perturbative component must be dominant. 
On the other hand, by assuming the existence of charm inside the proton 
we are consistently including the flavor exitation diagrams which are 
not considered in the LO calculation of Eq.~(\ref{eq1})-
(\ref{eq3})~\cite{halzen} (See Fig.~\ref{fig0b}). 
Note that flavor 
exitation diagrams are usually included in the pQCD calculation 
at NLO, but only for the perturbatively generated charm quarks. 
The non-perturbative charm sea must be taken into account through 
the recombination process. Moreover, it is difficult to account for the 
recombination of the spectator $c$-quark (see Fig.~\ref{fig0b}) when 
flavor exitation diagrams are included into a NLO pQCD calculation of charm production.

The recombination function $R_3$ is given by~\cite{hwa}
\begin{eqnarray}
R_3( \xu,& \xd &,\xc,\xf )=\alpha \frac{(\xu\xd )^{n_1}x_c^{n_2}}
{\xf^{n_1+n_2-1}} \nonumber \\
&\times& \delta \left(\xu+\xd+\xc-\xf \right)   
\label{eq9}
\end{eqnarray}
with $n_1=1$ and $n_2=5$~\cite{nos-epj} and $\alpha$ is a 
normalization constant. The same recombination 
function is used for $\alc$ inclusive production. 

In order to compare model predictions to experimental data, 
we have used 
\beq
\frac{dN^{\lc}}{d\xf} = N\left[\frac{d\sigma}{d\xf}_{Frag} 
+ r^{IC,Rec} \frac{d\sigma}{d\xf}^{\lc}_{Rec,IC} \right] \; ,
\label{eq10}
\eeq
and similarly for the $\alc$ differential cross section. 
In the equation above, $N$ is a global normalization constant 
which has been adequately fixed from experimental data. 
The $r^{IC,Rec}$ parameter was allowed to be 
different for the IC and R2C models, however, it was fixed to the 
same value for particle and antiparticle production within each model. 
The $\lc-\alc$ asymmetry as a function of $\xf$ was calculated according to 
Eq.~(\ref{eq0}) using the parametrization of Eq.~(\ref{eq10}) 
for the particle and antiparticle $\xf$ inclusive distributions. 

To control the size of each individual contribution to the 
total differential cross section, 
the calculated distributions for the fragmentation, IC and 
recombination processes for $\lc$ production were each normalized 
to unity. The recombination differential cross section for $\alc$ production 
was normalized by multiplying by $\sigma^{-1}$, where 
$\sigma=\int_0^1\frac{d\sigma}{d\xf}^{\lc}_{Rec}dx$ before normalization. 
Thus, the contribution of the light sea to the $\lc$ and $\alc$ production 
is the same. In this way, the coefficients $r^{IC}$ and $r^{Rec}$ give the relative 
size of the IC and Recombination component respectively, in comparison to the 
independent fragmentation process in the total cross section.

\begin{figure}[ht]
\centerline{\epsfig{figure=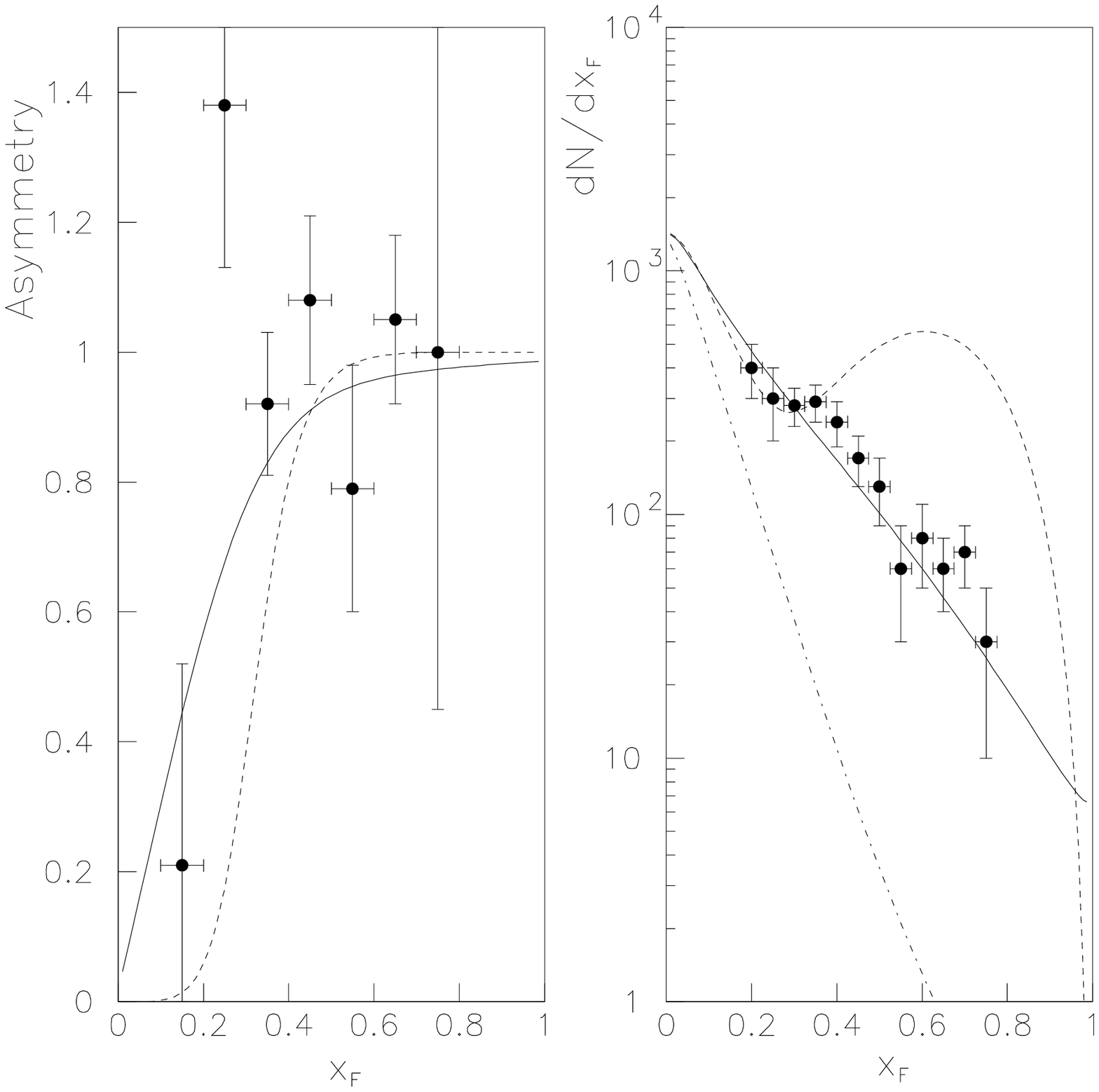,width=3.5in}}
\caption{Left: $\lc-\alc$ asymmetry in $pN$ 
interactions. Solid line is the prediction of the R2C 
model with $r^{Rec}=1.5$ and dashed line is the prediction of the 
IC model with $r^{IC}=0.4$. Right: $\lccc$ $x_F$ 
distribution in $pN$ interactions. Solid line is the prediction 
of the R2C model and dashed line is the prediction of the 
IC. Dot-dashed line shows the $\alc$ distribution 
as predicted by the R2C model. Experimental data are 
from Ref.~[3].}
\label{fig1}
\end{figure}

Prediction by the IC and R2C models are shown in 
Fig.~\ref{fig1} for both the $\lc-\alc$ asymmetry 
and the inclusive particle distribution as 
a function of $\xf$. Model predictions are also compared to 
the $p+N\rightarrow \lcc+X$ data from the SELEX 
Collaboration~\cite{selex}. 

Curves were obtained using the GRV-92~\cite{grv} parton distributions in 
nucleons in Eqs.~(\ref{eq3}), (\ref{eq7}) and (\ref{eq8}). 
The exponent $\gamma$ in the correlation function of Eq.~(\ref{eq8a}) 
was fixed to $\gamma=-0.1$~\cite{nos-prd} and the overall $Q^2$ scale was 
chosen as $Q^2=4m_c^2$ with $m_c=1.5$ GeV. In the Peterson fragmentation 
function we used $\epsilon=0.06$.
In order to fix the parameter $r^{IC,Rec}$ in both the IC and 
R2C models, we adjusted the curves to describe first the asymmetry. 
Once this parameter was fixed, we compared model predictions 
to experimental data on the $\xf$ particle distribution.

As can be seen in the figure, the IC model cannot describe 
simultaneously the $\lc-\alc$ asymmetry and the $\lccc$ $\xf$ 
distribution. On the other hand, the R2C model gives a qualitative and 
quantitatively good description of both, the  
asymmetry and the $\xf$ particle distribution.

\begin{figure}[ht]
\centerline{\epsfig{figure=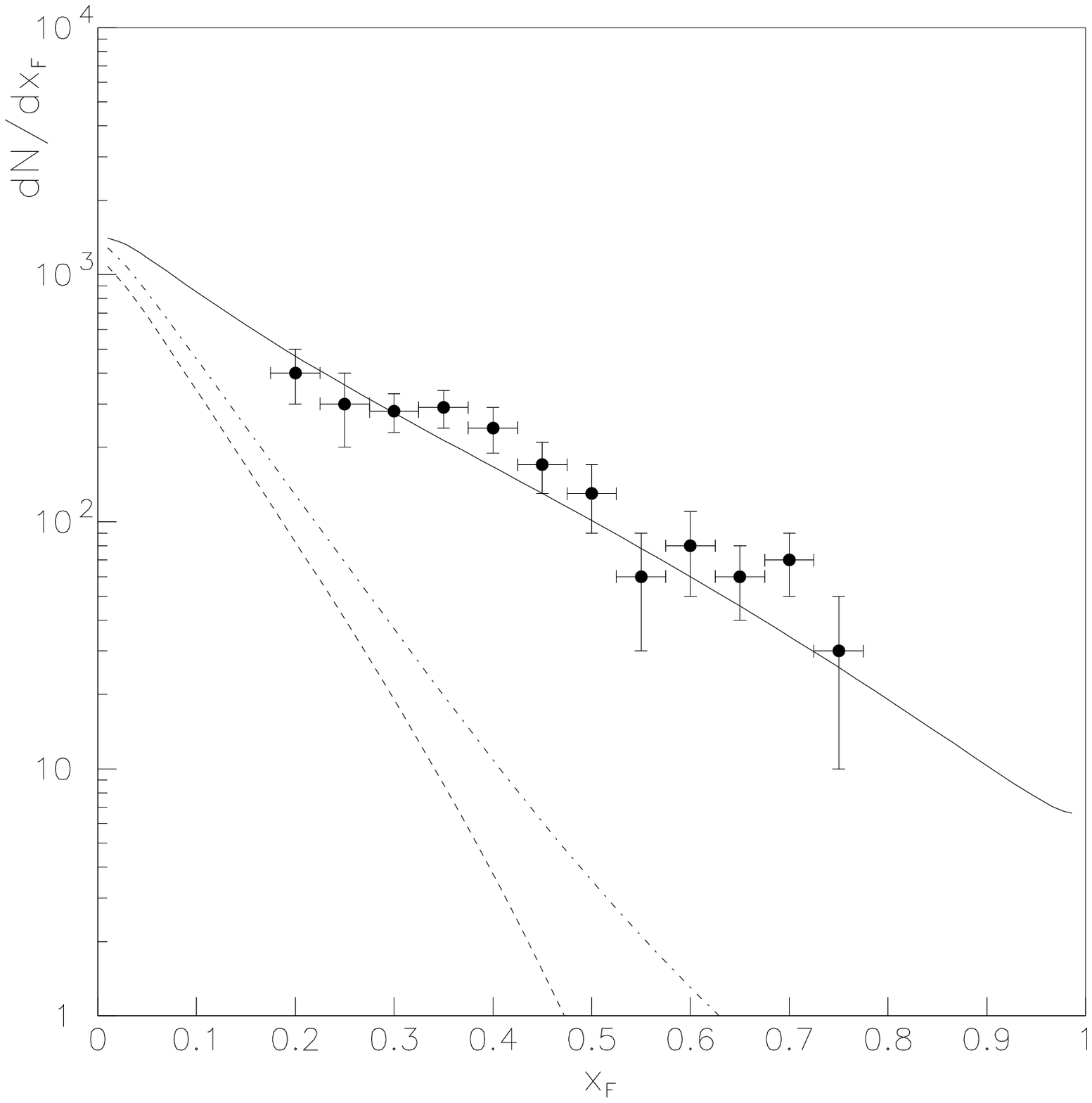,width=3.5in}}
\caption{$\lc$ (solid line) and $\alc$ (point dashed line) 
particle distributions predicted by the R2C model in 
$p-p$ interactions. Dashed line is the contribution to $\lc$ and 
$\alc$ production coming from the independent fragmentation. 
Experimental data are from Ref.~[3].}
\label{fig2}
\end{figure}

Furthermore, in order to fit the $\lccc$ $\xf$ distribution, 
the recombination part of Eq.~(\ref{eq1}) must be of the order of 
$1.5$ times bigger than the fragmentation contribution. This implies 
that the main mechanism in $\lccc$ production is recombination, 
even in the low $\xf$ region (see Fig.~\ref{fig2}). It is interesting 
to note also that recombination seems to be bigger than independent 
fragmentation also for $\alc$ production. 
The IC contribution to the proton structure, which is 
expected to be small, can only give a marginal contribution, and only 
at high $\xf$ ($\xf\rightarrow 1$) values, to the $\lcc$ production in 
proton-proton interactions. Thus, the recombination mechanism seems to be 
the principal contribution to the hadronization process, 
even more important than independent fragmentation of charm quarks.


\section*{Acknowledgments}


J.M and L.M. acknowledege J.A. Appel for useful comments and suggestions. 
The authors wish to acknowledge the Organizing Committee of 
SI\-LA\-FAE-III for the warm hospitality at the Conference. 
J.M is partially supported by COLCIENCIAS, the Colombian Agency for Science and 
Technology, under contract No. 242-99.

\end{document}